\documentclass[aps,pra,twocolumn,superscriptaddress]{revtex4-2}
\usepackage{amsmath}
\usepackage{amsfonts}
\usepackage{bm}
\usepackage{braket}
\usepackage{graphicx}
\usepackage{siunitx}
\usepackage{dsfont}
\usepackage[ruled,vlined]{algorithm2e}
\usepackage{xcolor}
\newcommand{\change}[1]{\textcolor{black}{{#1}}}
\usepackage{array}
\usepackage[T1]{fontenc}

\usepackage[colorlinks=true, allcolors=teal]{hyperref}

\usepackage{orcidlink}

\begin{document}
\title{Identification of quantum generative circuits with parallel quantum  neural network}
\author{Zheping Wu\,\orcidlink{0009-0008-1861-2685}}
\affiliation{%
School of Computer Science, Northwestern Polytechnical University, Xi’an 710129, China 
}

\author{Xiaopeng Huang}
\affiliation{%
school of cybersecurity, Northwestern Polytechnical University, Xi’an 710129, China 
}

\author{Hengyue Jia\,\orcidlink{0000-0002-5335-5329}}
\affiliation{%
School of Information, Central University of Finance and Economics, Beijing 100081, China
}

\author{Haobin Shi}
\affiliation{%
School of Computer Science, Northwestern Polytechnical University, Xi’an 710129, China 
}

\author{Wei-Wei Zhang\,\orcidlink{0000-0002-8164-9527}}
\email{Corresponding author: wei-wei.zhang@nwpu.edu.cn}

\affiliation{%
School of Computer Science, Northwestern Polytechnical University, Xi’an 710129, China 
}
\date{\today} 

\begin{abstract}
The rapid emergence of quantum technology has raised new challenges in distinguishing   various quantum circuits of similar functions. In this work, we propose parallel quantum embedding neural network (ParaQuanNet) for the efficient identification of quantum generative circuits via classifications of the corresponding output data.  Specifically, we generated W-like states with eight generative quantum  circuits realizing the generative quantum denoising diffusion probabilistic models (QDDPM). Our ParaQuanNet can classify these eight classes of generated quantum data with an accuracy of {$99.5\%$}, even though all of them are trained to generate the same types of quantum data.   With a novel design of parallel quantum embedding unit (PQEU) in our neural networks, our ParaQuanNet enables the quantum kernel circuit parallelly process all the receptive fields of quantum data, which empowers the quantum data processing efficiency. We also integrate the mutual unbiased  measurements into our ParaQuanNet and further improve its performance. 
 We apply our ParaQuanNet on the classification of classical data sets and demonstrate a good performance  of quantum neural networks on these tasks.  Our approach demonstrates good robustness to noisy data and the circuit-level noise \change{with a Python realization in calssical GPU}.  Our results highlight ParaQuanNet as a scalable and effective framework for quantum circuits identification,  contributing to the broader development of quantum machine intelligence.
\end{abstract}
\maketitle

\section{Introduction}
Quantum technology has attracted tremendous attention around the world and has been recognized as the height of the next information revolution. Within recent years, quantum technology has demonstrated breakthroughs in communications~\cite{liao2017satellite,cacciapuoti2019quantum,li2025microsatellite},  quantum computations~\cite{arute2019quantum,zhong2020quantum,krinner2022realizing,gao2025signal}, quantum simulations~\cite{o2016scalable,arguello2019analogue,PhysRevA.105.023513,king2025beyond,PhysRevResearch.6.043163,xvng-rylk} and quantum metrology~\cite{PhysRevLett.96.010401,PhysRevLett.112.080801,MONTENEGRO20251,Fadel_2025}. With the promising quantum advantages demonstrated, quantum technology is still at the stage of near term Noisy Intermediate-Scale Quantum era and there are still challenges in the field to find more practical applications for current quantum technologies~\cite{chen2023complexity,lau2022nisq}.

In the meantime, the development of artificial intelligence is also rising, showcasing its power in the emergence of intelligence~\cite{ho2020denoising,sanchez2018inverse,goodfellow2020generative}. Recent advances in large-scale deep learning—particularly transformer-based architectures have led to the development of models exhibiting emergent cognitive capabilities that were neither explicitly programmed nor anticipated at smaller scales~\cite{wei2022emergent,bowman2024eight}. These behaviors include in-context learning, abstract reasoning, and multimodal integration, suggesting the spontaneous organization of representational hierarchies akin to elements of natural cognition~\cite{bubeck2023sparks}. 

While the artificial intelligence is developing extremely fast, the corresponding computational power cost also grows vastly~\cite{sastry2024computing}. Attracted by the  demonstrated computational advantage of quantum technology and inspired by the development of artificial intelligence, scholars in the quantum field explore the power of quantum artificial intelligence. One direction aligning with this direction is the quantum variational algorithms for the optimization problems~\cite{kandala2017hardware,cerezo2021variational}, which are quantum-classical hybrid algorithms. The other direction is the exploration of the advantages of quantum artificial intelligence, such as the quantum ability in data mining~\cite{PhysRevResearch.6.043163}, classifications~\cite{nghiem2021unified,hur2022quantum}  and data generation~\cite{kolle2024quantum,wang2024quantum,minami2025generative}.  The quantum data generation varies in two ways, one is the generation of quantum representations of classical data, such as the MNIST data~\cite{kolle2024quantum,wang2024quantum,minami2025generative}. The other way is the generation of quantum states, such as quantum entangled states like GHZ and W states~\cite{zhang2025quantum,shaik2025quantum,parigi2024quantum}, and quantum multi-particle interaction systems~\cite{zhang2024generative,kwun2025mixed}. 
There are also efforts to develop the structure of quantum neural networks~\cite{he2026using,liu2026output}.

In this work, we focus on the aspect of copyright authentication in quantum artificial intelligence to advance the field of quantum AI further. We propose a method for the identification of quantum generative circuits with our novel  parallel quantum embedding
neural network (ParaQuanNet). Specifically, we identify the quantum generative circuits by the classifications of the quantum AI generated data from these quantum generative circuits. We introduce the mutual unbiased measurements into our ParaQuanNet to further improve the classification accuracy. 
Our results highlight ParaQuanNet as
a scalable and effective framework for quantum circuits identification, contributing to the broader
development of  quantum machine intelligence.
\begin{figure*}[th]
\centering
\includegraphics[width=480 pt, height=250 pt]{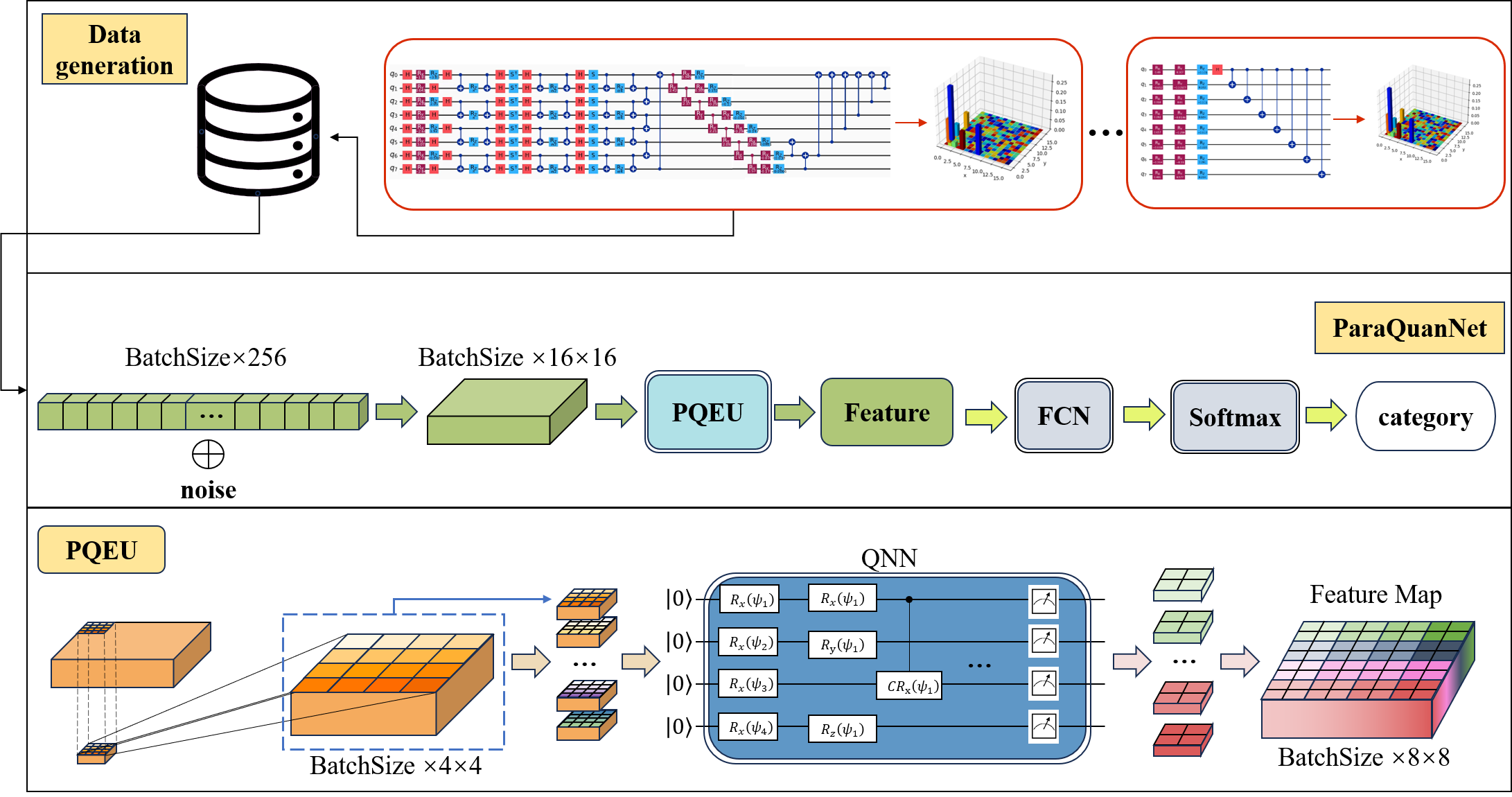} 
\caption{Sketch of our scheme for quantum identification of quantum AI-generated data. Top: the quantum data generated from various quantum generative circuits.   Middle: the workflow of our ParaQuanNet.  Bottom: the structure of the proposed  PQEU.}
\label{fig:scheme}
\end{figure*}
Our contributions are as follows: 
\begin{itemize}
    \item We forward the quantum artificial generation field for counterfeit tracking and copyright protection by identifying the quantum AI-generated data/ classifying the quantum generative circuits, which can also be used to classify classical data.
    \item We propose a parallel quantum embedding neural network (ParaQuanNet) enabling the implementation of the quantum convolution operations in parallel. 
    \item We design a parallel quantum embedding unit (PQEU) by utilizing a shared QNN as its kernel, and achieve 
parameter sharing advantages that significantly decrease model complexity while enhancing both training efficiency and generalization performance.
\item We integrate two types of mutual unbiased measurements into our ParaQuanNet, which further improves the classification accuracy.
  \item Our ParaQuanNet demonstrates the better robustness to quantum-specified  data noise and  circuit level noise compared with the traditional quantum convolution neural network.
\end{itemize}

\section{quantum AI-generated data identification with parallel quantum neural network}
This work aims to identify which quantum AI generative circuits are the quantum states  generated from and therefore to distinguish the quantum generative circuits. In this section, we propose a parallel quantum embedding neural network (ParaQuanNet) architecture for the efficient quantum generative circuits identification.  Specifically, we will introduce the generative circuits we used as demonstration in this works,  the construction of  ParaQuanNet and its core part of  parallel quantum embedding network, the novel integration of mutual unbiased measurement in our quantum ParaQuanNEt, and our main results of quantum generative circuits identificaiton. 
\begin{figure}[t]
\centering
\includegraphics[width=0.98\columnwidth]{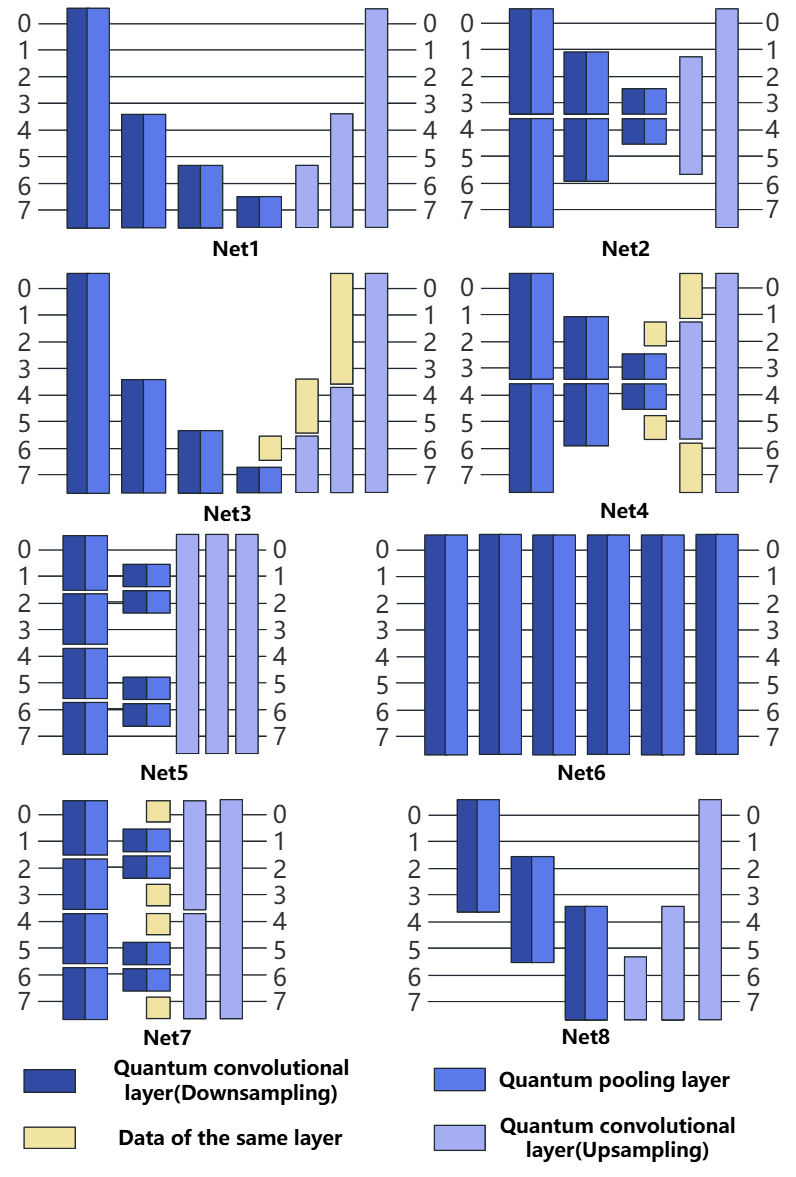} 
\caption{The eight structures of quantum W-like states generation circuits used for our data generation.}
\label{fig:eight-UNet}
\end{figure}

The architecture of our Parallel  quantum neural network is shown in Fig.~\ref{fig:scheme}, and it consists of three parts, the data generated process, the quantum neural networks and the measurements. In the data generated process, we collect the quantum data from various quantum generative circuits which are trained to generate the same types of data. In our demonstration, we identify 8 types of quantum generative circuits. The quantum neural networks we design include a parallel quantum embedding network  and enable the parallel quantum convolution. We integrate the mutual unbiased measurements into our quantum neural networks to improve the performance further.

\subsection{Quantum generative circuits}

The data in our proposal is generated with various quantum generative circuits which are trained by following a quantum  denoising diffusion probabilistic process~\cite{ho2020denoising,zhang2025quantum,shaik2025quantum,parigi2024quantum}.  Specifically, quantum diffusion model  extends classical denoising diffusion probabilistic models (DDPMs) into the domain of quantum states by replacing Gaussian noise with quantum channels and learning to reverse this noise process using quantum neural networks. A practical  effective architecture for this task is the quantum U-Net in Ref.~\cite{zhang2025quantum}, a quantum analogue of the classical U-Net that uses hierarchical encoding and decoding with skip connections to preserve multiscale structure in quantum data. In a quantum DDPM with a quantum U-Net, the forward process gradually perturbs an initial quantum state through controlled CPTP noise channels, while the quantum U-Net learns the reverse transitions by predicting and removing the injected quantum noise at each diffusion step. This combination provides a powerful framework for quantum state reconstruction, generation, and noise mitigation, leveraging the expressive capacity of quantum circuits to capture both local and global correlations that arise in many-body quantum systems.

In this work, we design various neural network architectures and train these neural networks to generate the same kind of quantum data, i.e. the previously defined the generated W-like state, which is defined as follows,
\begin{equation}
\begin{split}
\widetilde{\text{W}}=&\alpha_1|10000000\rangle+\alpha_2|01000000\rangle\\+&\alpha_3|00100000\rangle+\alpha_4|00010000\rangle\\+&\alpha_5|00001000\rangle+\alpha_6|00000100\rangle\\+&\alpha_7|00000010\rangle+\alpha_8|00000001\rangle
\end{split}  
\label{eq:w-1}
\end{equation}  
with 
$\alpha_{i}, i\in[1,...,8]$ as complex numbers and  $\sum_i|\alpha_i|^2=1$. We define the success rate of generations is \begin{equation}
\begin{split}
P_\textbf{succ}^{\text{W}}=&P_\text{10000000}+P_\text{01000000}\\+&P_\text{00100000}+P_\text{00010000}\\+&P_\text{00001000}+P_\text{00000100}\\+&P_\text{00000010}+P_\text{00000001}
\end{split}
\end{equation}
Here $P_\text{00000000}$, $P_\text{111111111}$, $P_\text{10000000}$, $P_\text{01000000}$, $P_\text{00100000}$, $P_\text{00010000}$, $P_\text{00001000}$, $P_\text{00000100}$, $P_\text{00000010}$ and  $P_\text{00000001}$ are the probabilities of the generated states project to the corresponding computation basis \{$\ket{00000000}$, $\ket{11111111}$, $\ket{00000001}$, $\ket{00000010}$, 
$\ket{00000100}$, 
 $\ket{00001000}$, $\ket{00010000}$, $\ket{00100000}$, $\ket{01000000}$, $\ket{10000000}$\}.

Specifically, in our demonstrations we design 8 various generative circuits as shown in Fig.~\ref{fig:eight-UNet}, where a convolution layer applies a single quasi-local unitary ($U_i$) in a translation invariant manner for finite depth, inspired by Ref.~\cite{cong2019quantum}.  For pooling, a fraction of qubits are mathematically traced out temporally, and their states are stored and reused in the later coming layers.  
Therefore, the nonlinearities in our architectures arise from tracing out partial degrees of freedom. With these 8 well trained generative circuits, we generate { 2000} data from each quantum generative circuits, and the success rate of the generations for each of these circuits are above {0.95}. In our work, we will use our proposed parallel quantum embedding neural network to identify which of these generative circuits are the quantum data generated from.

\subsection{Parallel quantum embedding network}
In this work, we propose a parallel quantum embedding network for the efficient data processing in quantum neural networks (ParaQuanNet).  Our novel multi-subgraph parallel quantum computing unit (PQEU) is designed in our ParaQuanNet as shown at the bottom of Fig.~\ref{fig:scheme}, under the inspiration of the high efficiency of GPU parallel computing, i.e., SIMD (Single Instruction Multiple Data) architecture~\cite{wang2013multiple}. By leveraging parameter sharing, localized connectivity, and efficient lattice-based data processing, PQEU achieves stronger feature representation capabilities with fewer parameters and significantly accelerates the computational efficiency of quantum convolutional neural networks (QCNNs). 
Specifically, our ParaQuanNet consists of three components: data preprocessing, feature extraction via PQEU, feature fusion, and classification. Quantum-generated data, inherently represented as complex-valued vectors, and mapped into distinct 2D tensors. The restructured data is processed through independent PQEU blocks, enabling simultaneous extraction of localized quantum features while maintaining parameter efficiency through weight sharing. Extracted features are fused via entanglement-inspired aggregation before being classified using a fully connected network. A softmax-activated output layer generates probabilistic predictions for classifying quantum states.

During the training process, our data is reorganized in the following way. Firstly, the input quantum data $\mathbf{z} \in \mathbf{C}^{ n \times 256}$  are  reshaped into the size of $\mathcal{T} \in \mathbf{R}^{n \times 16 \times 16}$, where $n$ represents the batchsize, and the process is classical implemented in our simulations. Then, the processed  data sets are fed into our shared-parameter PQEU modules for the quantum feature extraction:  
    \[
    \mathbf{f}_* = \text{PQEU}(\mathcal{T}; U(\theta)),
    \]  
with $U(\theta)$ representing the quantum neural network in our PQEU. 
In the last step, the fused features are sent through a fully convolution network followed by Softmax activation, and output the final class identification category of the data.

\begin{figure}[t]
\centering
\includegraphics[width=0.98\columnwidth]{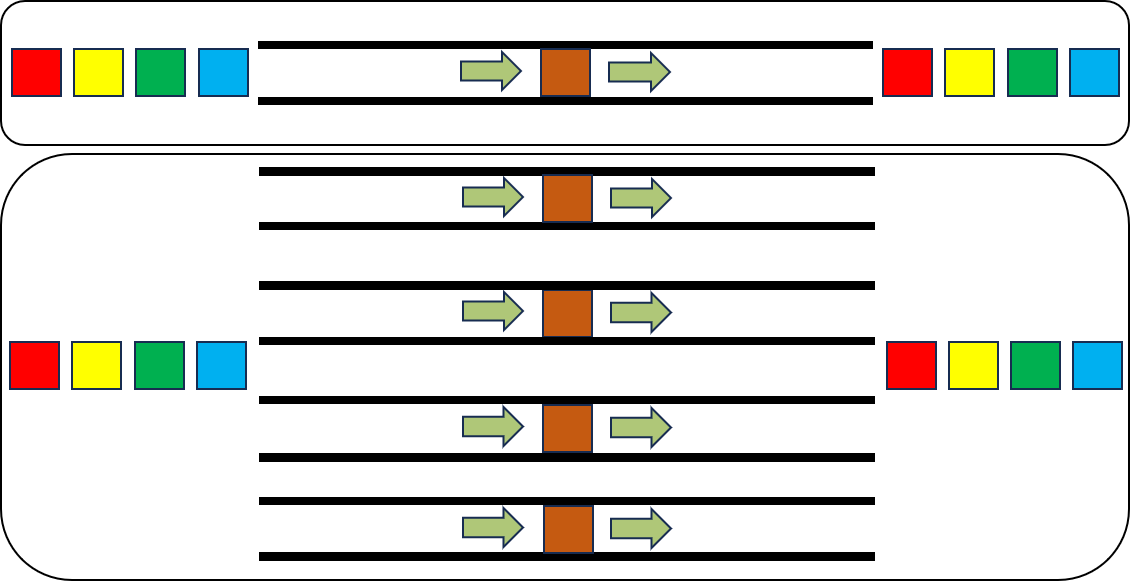} \caption{Top: Convention quantum neural network data flow. Bottom: the sketch of our \change{PQEU} process data flow.}
\label{fig:Q-SIMD}
\end{figure}
The key innovation in our ParaQuanNet is the parallel quantum embedding unit (PQEU), which upgrades the traditional quantum convolution network by enabling the processing of multiple patches simultaneously and achieves both spatial feature extraction and computational acceleration as shown in Fig.~\ref{fig:Q-SIMD}. In our simulations, we realize it in GPU. While applying in quantum NISQ devices, it requires multi copies of trained PQEU for the process. The mechanism of our PQEU was invented under the inspiration of the SIMD (Single Instruction Multiple Data) architecture in classical parallel computing. The workflow of PQEU is shown at the bottom of Fig.~\ref{fig:scheme}. The structure of our QNN consists of the single qubit gates $R_\text{x},R_\text{y},R_\text{z}$,  SX gate, Hadamard gates, and the two-qubit CNOT gates and controlled rotation gates $CR_\text{x},CR_\text{y},CR_\text{z}$.  In the workflow of our 
PQEU, the first part is a data griding process, which converts the data with the size of $\text{BatchSize}\times16\times16$ into $16\times\text{BatchSize}$ pieces of the data with a size of $4\times 4$. These  $16\times\text{BatchSize}$ pieces of data are further vertically stacked  for our 4-qubit convolution kernel quantum neural network (QNN). The output of the QNN is a vertically stacked feature with a size of $2\times 2$, which is further fused into a feature map with a size of $8\times 8$.

We summarize to emphasize the two critical innovations of our PQEU as follows.
\begin{itemize}
    \item Batch Griding: Instead of processing patches sequentially, our PQEU converts a batch of `patches' into a 4-qubit system, enabling simultaneous processing of multiple patches in one round of the QNN.
    \item Shared Parameter Convolution: All patches within a batch share identical parameterized quantum gates, reducing the number of circuit free parameters while maintaining spatial feature locality.
\end{itemize}

\begin{figure}[th]\centering
\includegraphics[width=0.95\columnwidth]{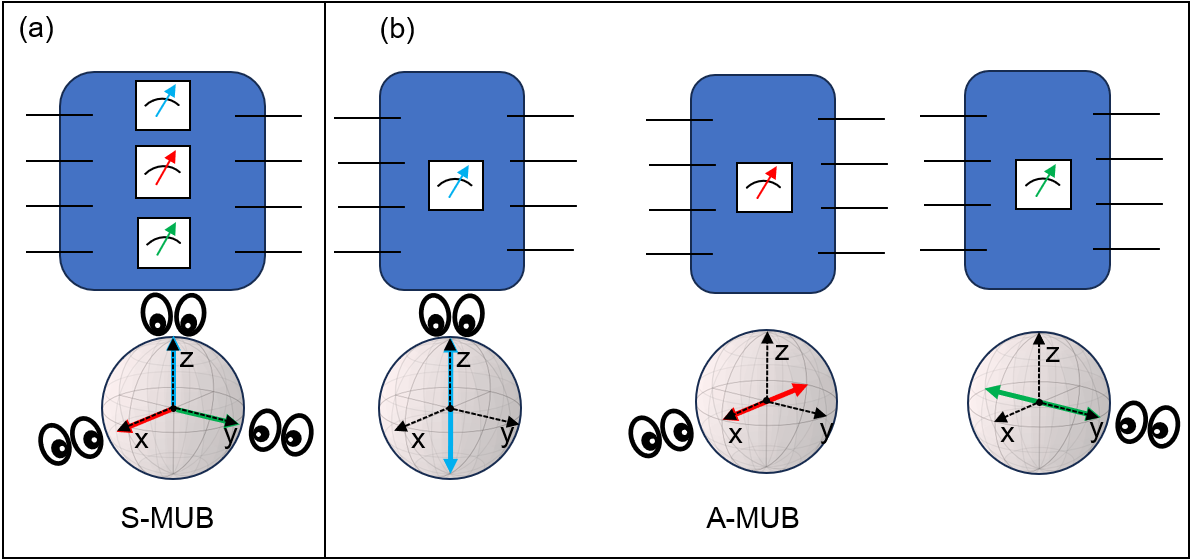}
\caption{Sketch of our two proposed measurement modes: (a) S-MUB and (b) A-MUB, where the eyes represent the quantum observations/measurements.} 
\label{fig:MUB}
\end{figure}
\subsection{Mutual unbiased measurements empowered ParaQuanNet}

In this section, we unleash quantum advantage for neural networks by introducing mutual unbiased measurement in two modes and experimentally demonstrate the great improvements of our framework in classification accuracy by the average improvement of { 18.9\%} compared with the model with ordinary computational measurements.

Mutually unbiased  bases (MUB) measurements  are essential for quantum state tomography, as they use measurement bases that are maximally incompatible, meaning knowledge of a quantum state in one base reveals no information about its state in the others. By combining outcomes from these bases researchers can reconstruct a quantum state’s full density matrix with optimal efficiency, minimizing redundancy and ensuring complete characterization. This approach is critical for verifying quantum systems in experiments, such as validating entangled states or calibrating qubits, as it maximizes information extraction while reducing measurement overhead~\cite{bent2015experimental,zhang2018adaptive,designolle2019quantifying,tavakoli2021mutually,farkas2023mutually}.  Mathematically, a pair of MUBs as two orthonormal bases on a $d$-dimensional Hilbert space $\mathbb{C}^d$, namely, $\{\ket{e_j}\}_\text{j=1}^d$ and 
$\{\ket{f_k}\}_\text{k=1}^d$, with the property that $| \left< e_j|f_k\right>|^2=\frac{1}{d}$ for all $j$ and $k$. 

Specifically, we introduce mutually unbiased  bases measurements in our ParaQuanNet to fully release quantum advantage and empower the  learning efficiency and  precision. For a single qubit with 2-dimensional Hilbert space, the three mutually unbiased  bases are the  Pauli X, Y, Z Bases. The measurement operators for Pauli X, Y, Z Bases are $\{\hat{\sigma}_x, \hat{\sigma}_y,\hat{\sigma}_z\}$ and the corresponding measurement bases are 
$\{|0\rangle, |1\rangle\}$ (computational  bases), $\{|+\rangle=\frac{1}{\sqrt{2}}(|0\rangle + |1\rangle), |-\rangle=\frac{1}{\sqrt{2}}(|0\rangle - |1\rangle)\}$, $\{|R\rangle=\frac{1}{\sqrt{2}}(|0\rangle + i|1\rangle), |L\rangle=\frac{1}{\sqrt{2}}(|0\rangle - i|1\rangle)\}$.  The key property is that the overlap between any two states from different bases is $|\langle\psi_i|\phi_j\rangle|=\frac{1}{\sqrt{2}}$, ensuring maximal uncertainty.  
To fully reconstruct a qubit’s density matrix $\rho$, we need its projections onto these three bases, 
$ \rho = \frac{1}{2}\left(I + \langle X \rangle X + \langle Y \rangle Y + \langle Z \rangle Z\right)$, 
where $\langle \hat{\sigma}\rangle_x, \langle \sigma \rangle_y, \langle \sigma \rangle_z$ are expectation values obtained by measuring in the X, Y, and Z bases. Each  bases provides complementary information, Z bases Reveals populations (diagonal elements of \(\rho\)), X/Y bases Reveal coherence (off-diagonal elements). The completeness and minimal redundancy for learning the information of a qubit are guaranteed, since the three Pauli bases span the space of 2×2 Hermitian matrices, ensuring no missing information, and each measurement contributes unique, non-overlapping data.

For the measurement in our ParaQuanNet for classifications, we propose two frameworks of MUB measurement, namely, simultaneous MUB measurements (S-MUB) and alternating MUB measurements (A-MUB) as shown in  Fig.~\ref{fig:MUB}. In S-MUB, the qubits in QCNN are measured in the three MUB  simultaneously as shown in  Fig.~\ref{fig:MUB} (a), the averaged measurement outcomes are used for the QCNN training. 
In A-SUB, the qubits in the QCNN are measured alternately in sequence by three MUBs (the timely order in our experiments are Paul Z, Pauli X, Pauli Y) as shown in  Fig.~\ref{fig:MUB} (b). Operationally, due to the disturbance from quantum measurement, quantum measurements are applied on multiple copies of the quantum states, and then we extract the statistical results from these sample measurements.  Compared with Pauli-Z measurements, S-MUB and A-MUB is more reflective quantum properties of quantum states and extract more information from measurements. Therefore, it will be shown in our simulations that  the S-MUB and A-MUB will give higher classification accuracy compared with the Pauli-Z.

\subsection{Quantum generative circuits identificaiton with our ParaQuanNet}

In this study, we employ ParaQuanNet to characterize quantum generative circuits by classifying the data produced by those circuits. Through this classification-based approach, our framework effectively links the structure of each quantum generative model to the statistical properties of its output, enabling a systematic identification of the underlying generative mechanisms. The pseudocode of our algorithm is shown in Alg.~\ref{alg:QuanvSingle}. All of our results are simulated in classical computer.
\begin{figure}[t]
\centering
\includegraphics[width=0.94\columnwidth]{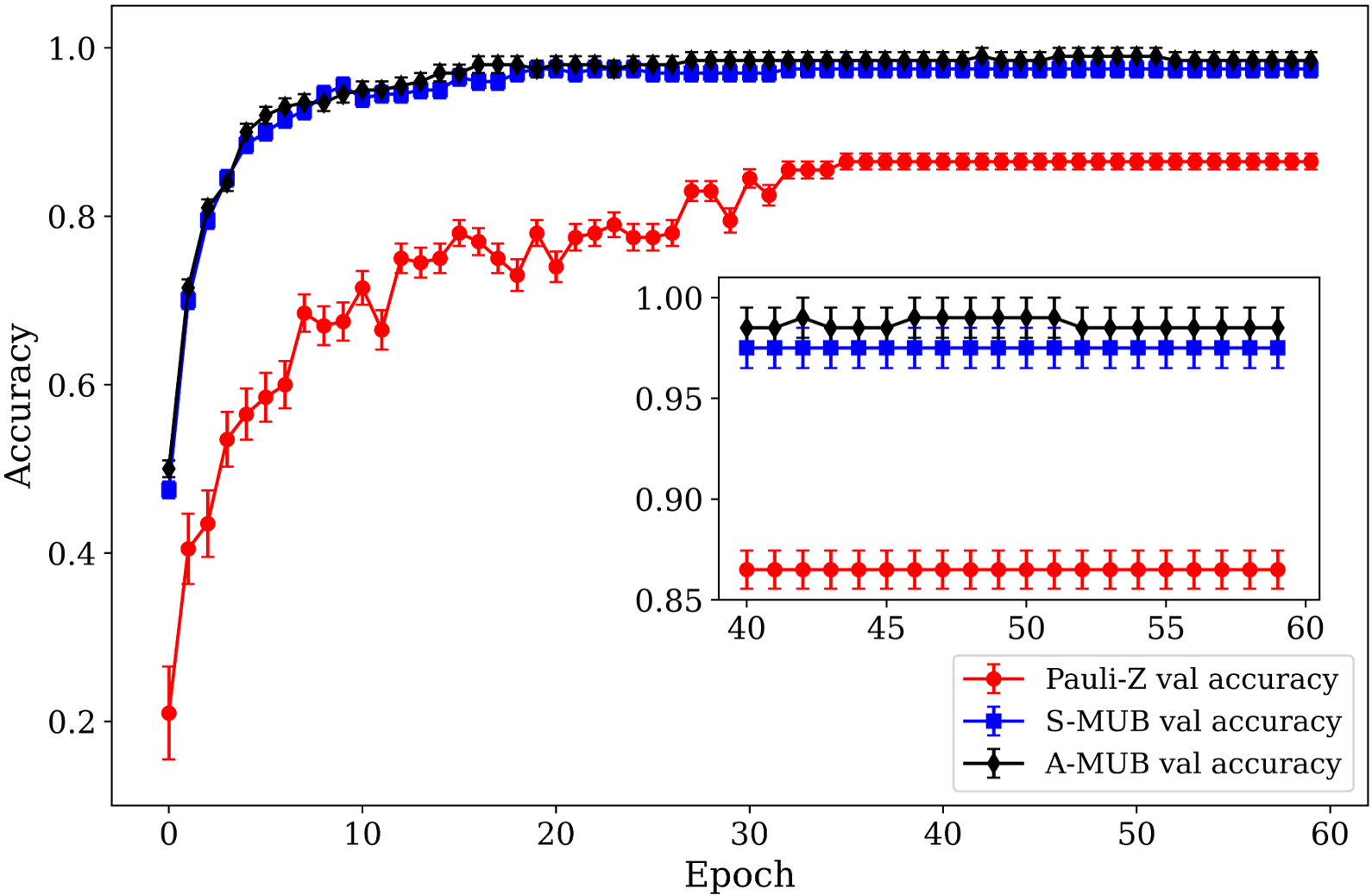} 
\caption{The classification accuracy for the eight classes of quantum data.}
\label{fig:acc-loss}
\end{figure}

\begin{algorithm}[t]
\caption{Quanvolution-based hybrid quantum--classical neural network}
\label{alg:QuanvSingle}
\SetAlgoLined
\DontPrintSemicolon

Initialize quantum device \(\mathcal{Q}\) with \(n_{\mathrm{wires}}\) qubits\;
Initialize quantum convolution kernel (parameterized quantum circuit) \(\mathrm{QLayer}\)\;

Initialize a linear layer \(W: \mathbb{R}^{64} \rightarrow \mathbb{R}^{8}\)\;

\For{each input batch \( x \) with batch size \( B \)}{

    Reshape each sample \( x \) into a \(16 \times 16\) grid\;
    Initialize tensor \(\mathrm{patches} \in \mathbb{R}^{B \times 4 \times 4 \times 16}\)\;

    \For{each patch row index \( c \)}{
        \For{each patch column index \( r \)}{
            Extract a \(4 \times 4\) patch from rows \([c, c+4)\) and columns \([r, r+4)\)\;
            Flatten each patch into a 16-dimensional vector and store in \(\mathrm{patches}\)\;
        }
    }
    Reshape \(\mathrm{patches}\) into a batch of patch vectors:
    \(\mathrm{patches} \in \mathbb{R}^{(B \cdot 16) \times 16}\)\;

    Initialize the quantum device \(\mathcal{Q}\) by encode the simulated quantum statevector into the quantum device via amplitude encoding.\;

    Apply \(\mathrm{QLayer}\) on \(\mathcal{Q}\)\;

    \eIf{measurement mode = 1}{
        Measure all qubits using \(M^Z\) and obtain feature matrix \(\mathrm{featureResult}\)\;
    }{
        \eIf{measurement mode  = 2}{
            Measure all qubits using \(M^X\) and obtain \(\mathrm{featureResult}\)\;
        }{
            Measure all qubits using \(M^Y\) and obtain \(\mathrm{featureResult}\)\;
        }
    }

    Reshape measurement results into feature vectors:
    \(\mathrm{featureResult} \in \mathbb{R}^{B \times 64}\)\;
    \( y = W \mathrm{featureResult} \in \mathbb{R}^{B \times 8} \)\;
    Compute final prediction via log-softmax:
    \( \hat{y} = \mathrm{softmax}(y) \)\;

}
\end{algorithm}

For the eight groups of W-like states generated the eight quantum generative circuits in Fig.~\ref{fig:eight-UNet}, the classification accuracy reaches {$99.5\%$} with A-MUB or S-MUB and the loss converges well, as shown in Fig.~\ref{fig:acc-loss}, much higher than the traditional Pauli Z measurements.  Empowered with PQEU, our ParaQuanNet demonstrates its superiority in efficiency 
compared with the existing quantum-classical hybrid models, as shown in Table~\ref{table:effiency}. Parameter efficiency of our ParaQuanNet  is enhanced dramatically compared to standard Quantum CNNs. Specifically, the number of parameters in our  ParaQuanNet is around $29\%$ of the number of parameters in QCNN, and the amount of samples being processed per second in our ParaQuanNet is 24 times of  that in QCNN.

The high classification accuracy demonstrate that even all the 8 generative circuits are trained to generate the same type of quantum data, but the corresponding quantum DDPM processes learned during the neural network training varies.  Our PareQuanNet are effective method to learn the difference of these various quantum DDPM processes represented by the trained  generative circuits.

\begin{table}[th]
\caption{Quantum generative circuits classifications}
\label{table:effiency}
\begin{ruledtabular}
\begin{tabular}{ccc}
Method & QCNN & ParaQuanNet  \\\hline
Average Accuracy & 95$\%$ &  99.5$\%$   \\\hline
$\#$ of Parameters & 2194 &  637  \\\hline
Samples processed/s & 61 &  1011   \\
\end{tabular}
\end{ruledtabular}
\end{table}

\begin{figure}[t]
\centering
\includegraphics[width=0.98\columnwidth]{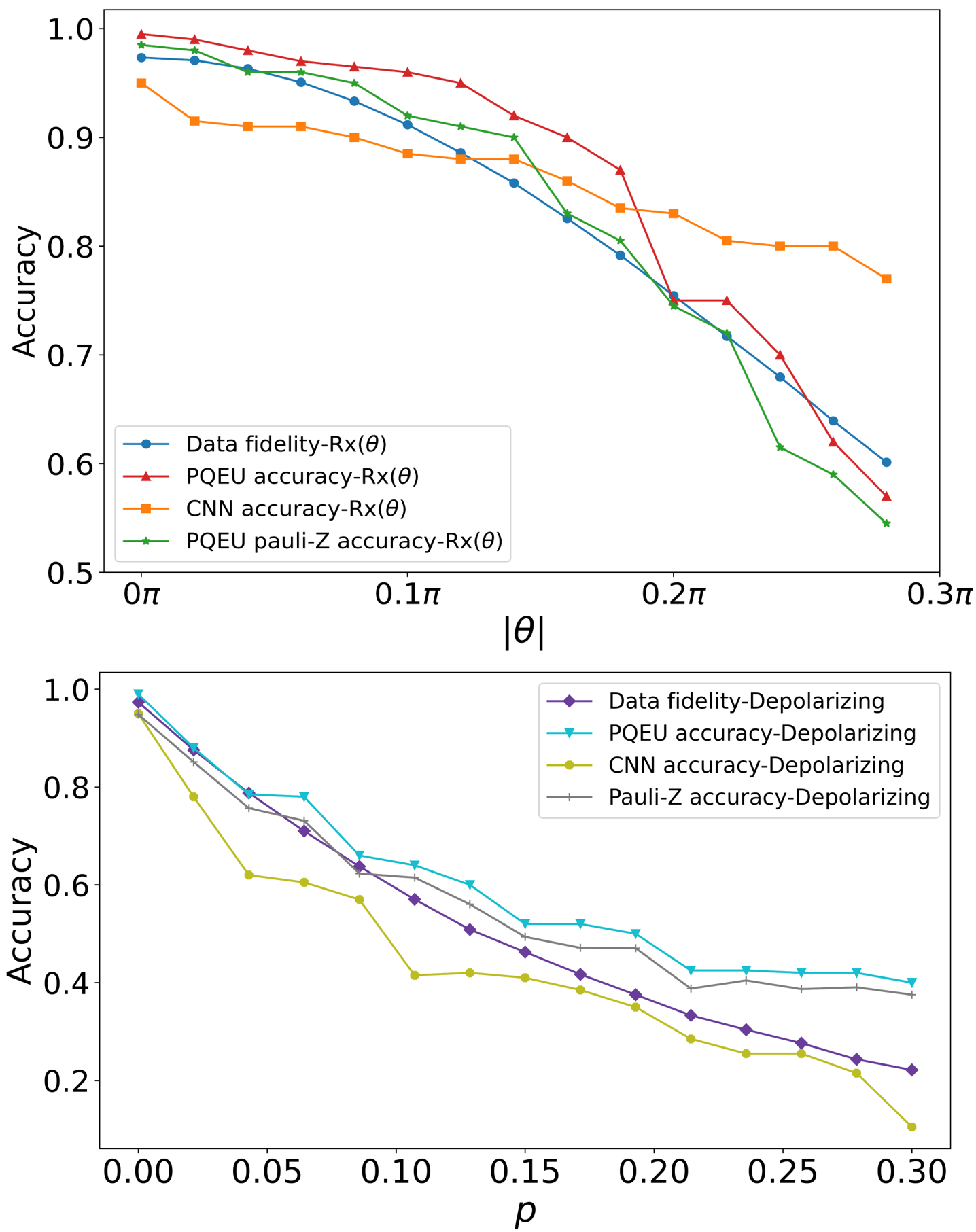} 
\caption{The robustness of our ParaQuanNet to the $R_\text{x}(\theta)$ noise (Top)  and depolarizing noise $\epsilon(p)$ (Bottom).}
\label{fig:noist-robustness}
\end{figure}

To study the robustness of our ParaQuanNet, we consider two types of data noise, the single-qubit $R_\text{x}(\theta)$ noise and the depolarizing noise $\epsilon(\rho)=(1-p)\rho+\frac{p}{3}(\hat{X}\rho \hat{X}+\hat{Y}\rho \hat{Y}+\hat{Z}\rho \hat{Z})$, with $\hat{X}, \hat{Y}, \hat{Z}$  representing the Pauli  operators. We artificially add these two types of noise to all the training data, and using these noised data to train our neural networks.
And by adding the various level noise into our generated quantum data, we test our ParaQuanNet on its classification accuracy.  As shown in Fig.~\ref{fig:noist-robustness}, we see that both the classification accuracy of our ParaQuanNet and the data fidelity drop as the noise level increases.

Specifically, our ParaQuanNet performs well (with accuracy above 90\%) while the $R_\text{x}(\theta)$ noise level is within $|\theta|<0.1\pi$ and the depolarizing noise level is within $p<0.02$. In contrast, the CNN method drops faster with the increase of noise level within the range where the classification accuracy is higher than 90\%.  The robustness of our method to the noise is consistent with the fidelity of the noise data still belonging to the W-like state group as shown in Fig.~\ref{fig:noist-robustness}. As the fidelity of the noise data belonging to the W-like states drops, the identification accuracy of our method also drops, which is reasonable. We would like to emphasize that our ParaQuanNet is not only efficient in terms of data processing speed and the lower number of internet parameters, but also in the better robustness to the quantum-specialized noise types, such as the depolarizing noise. Specifically, with the increase of noise level, the classification accuracy of our ParaQuanNet drops much slower than CNN.

We further study the robustness of our approach by considering  three types of realistic noisy cases, i.e. two-qubit gate fidelity, finite-shot measurement noise and readout errors. We have updated our manuscript accordingly.
	\begin{figure*}[t]
		\centering
	(a)\includegraphics[width=0.45\textwidth]{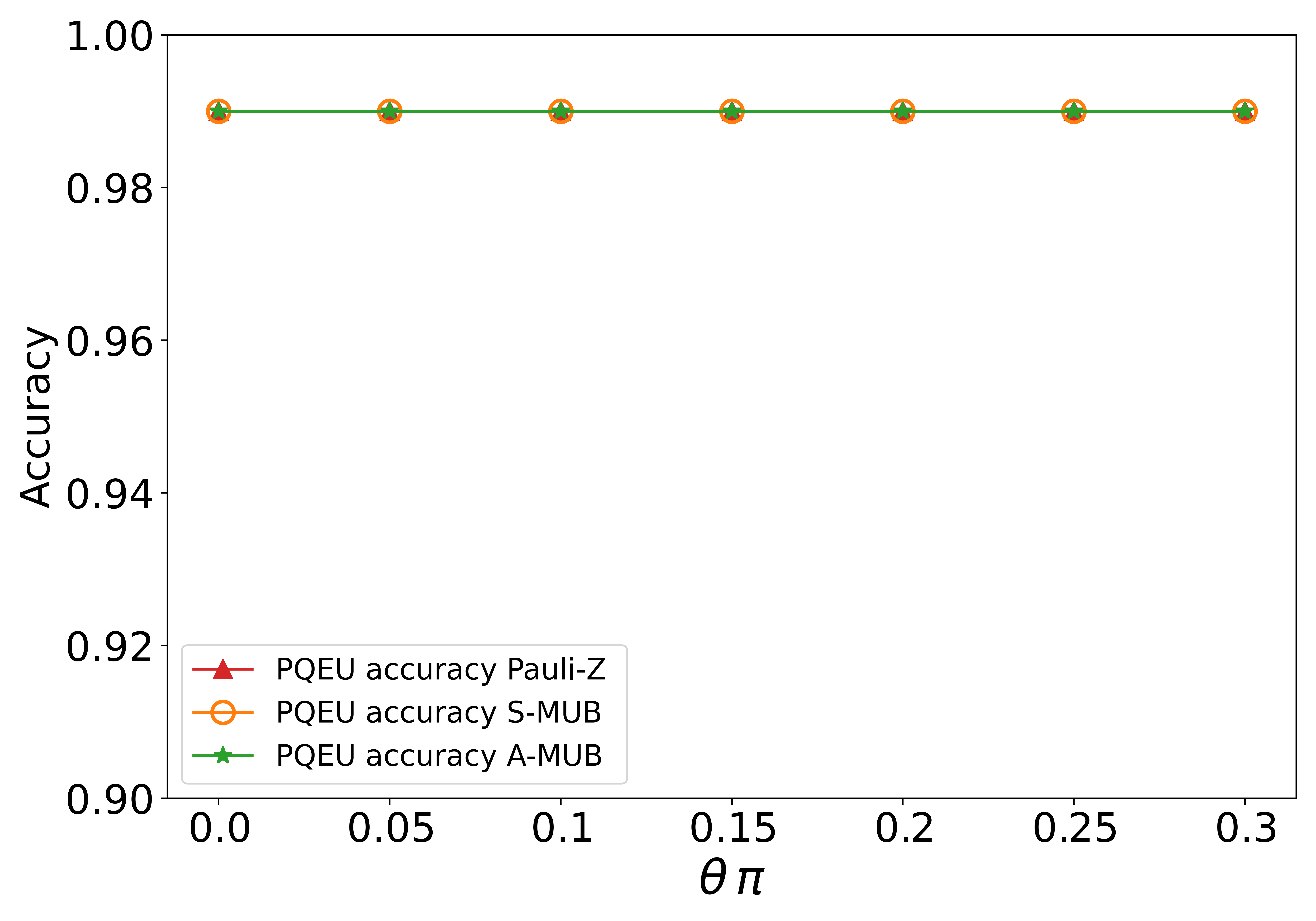}	(b)\includegraphics[width=0.45\textwidth]{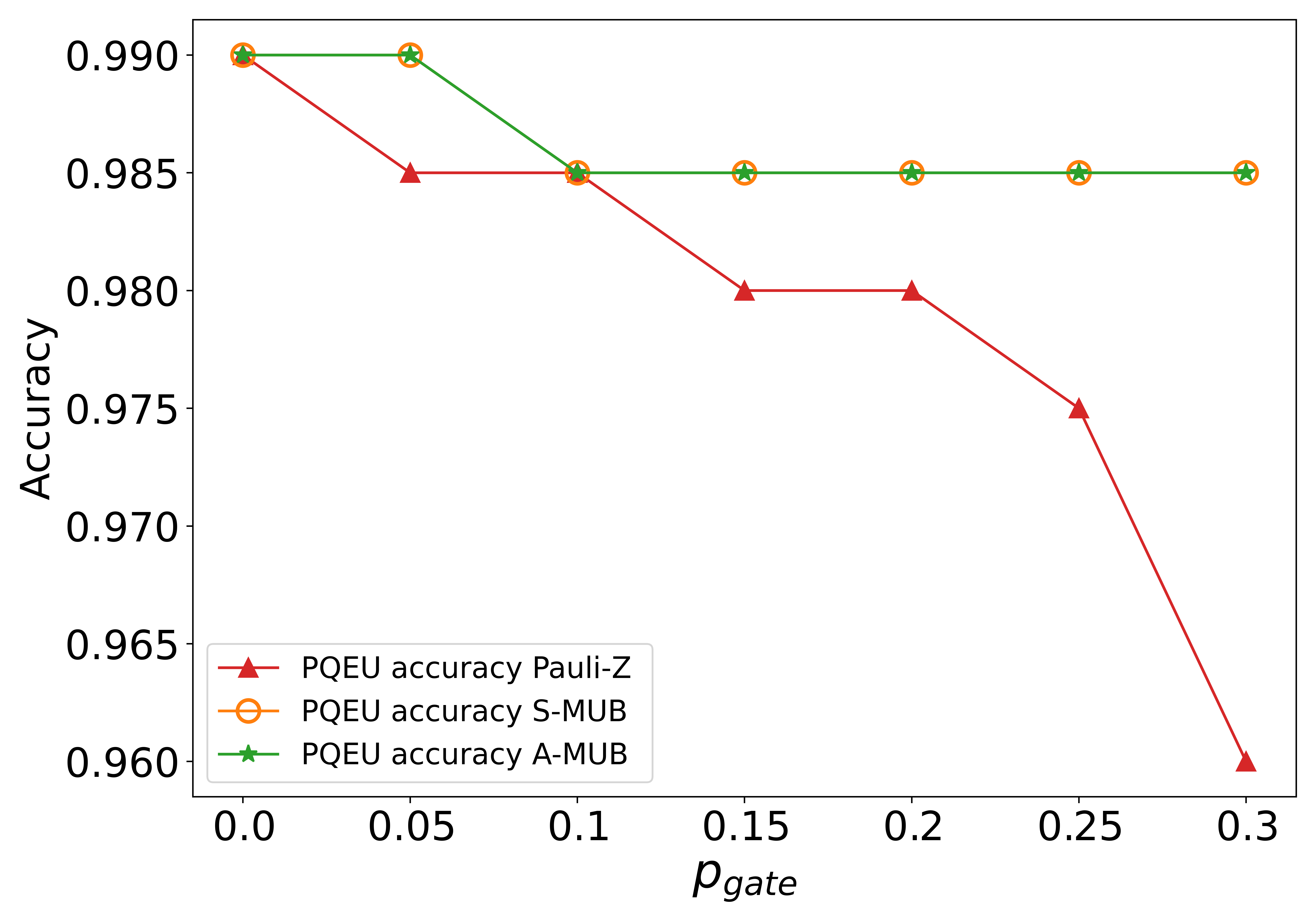}
		(c)\includegraphics[width=0.99\textwidth]{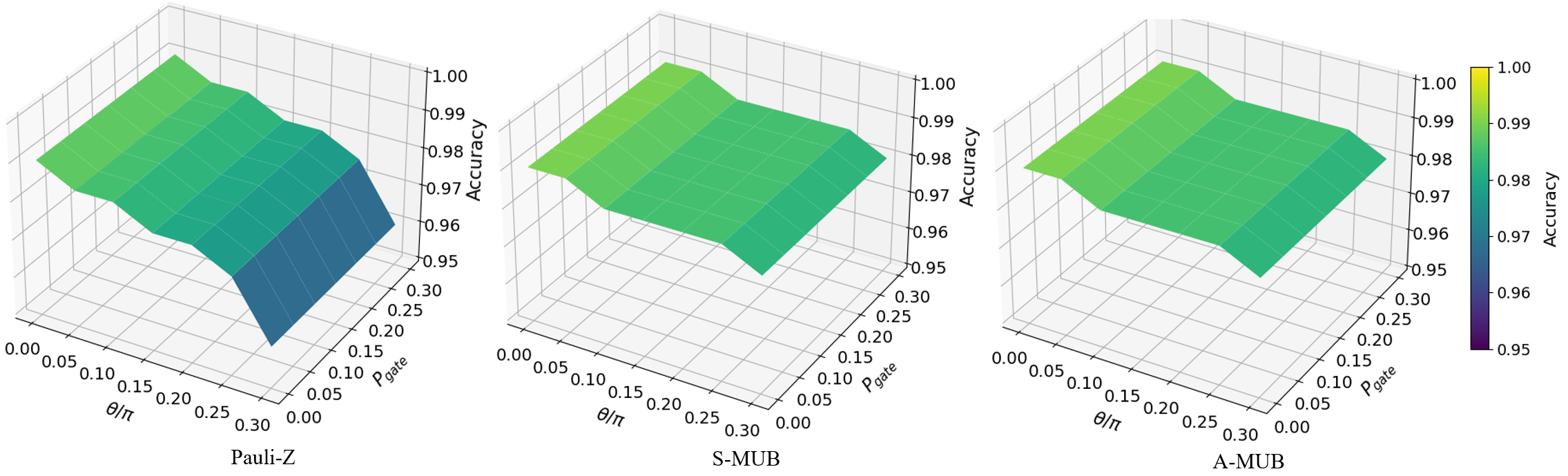}
	\caption{Classification accuracy under (a) single-qubit gate noise, (b) two-qubit gate noise and (c) both single- and two-qubit gate noise under Pauli-A (left), S-MUB (center), A-MUB (right) measurement strategies.}
	\label{fig:gate_noise}
	\end{figure*}

\textbf{Our approach performance considering single-qubit and two-qubit gate fidelity.} The gate fidelity is one of the dominating issues in quantum computation. Our circuit is composite of single-qubit and two-qubit gates. Here we consider imperfect single-qubit and two qubit gates. Imperfect single-qubit  gates are modeled using an equivalent parameterized $R_x(\theta)$ operator in the end of each qubit in our PQEU kernel. Imperfect two-qubit entangling gates are modeled using a two-qubit depolarizing channel applied after each two-qubit operation (e.g., CNOT and controlled rotations):
	\begin{equation}
	\mathcal{E}^{(2)}_{p}(\rho)
		=
		(1 - p_\text{gate})\rho
		+
		p_\text{gate} I_4,
	\end{equation}
	where $p_\text{gate}$ denotes the depolarizing probability and $I_4$ is the $4\times4$ identity matrix.
	In simulation, this channel is implemented via a Pauli-twirled approximation,
	\begin{equation}
		\rho \;\mapsto\;
		(1 - p_\text{gate})\rho
		+
		\frac{p_\text{gate}}{15}
		\sum_{\substack{P,Q \in \{I,X,Y,Z\} \\ (P,Q)\neq(I,I)}}
		(P \otimes Q)\rho(P \otimes Q),
	\end{equation}
	which provides a standard statevector-compatible approximation of two-qubit depolarizing noise.

As shown in Fig.~\ref{fig:gate_noise}, we implement our simulations under various gate fidelity level. 
In Fig.~\ref{fig:gate_noise} (a), we vary the parameter $\theta\in[0,0.3\pi]$, we observe the robustness of our method to this single-qubit gate noise. 
We vary  $p_\text{gate}$ from  to a realistic high value of 0.3 in Fig.~\ref{fig:gate_noise} (b). It is shown that our approach has a good robustness to the gate fidelity, the accuracy can reach 0.985 even at the depolarizing error $p_\text{gate}=0.3$. We also witness that our S-MUB and A-MUB method outperform the standard the Pauli-Z measurement method, demonstrate better robustness against the gate fidelity. 
In Fig.~\ref{fig:gate_noise} (c), we show the performance of our method while considering both single-qubit and two-qubit gate noise, where left is Pauli-Z measuremnt case, center is the S-MUB case, and right is the A-MUB case.
	\begin{figure}[h]
		\centering
	\includegraphics[width=0.95\columnwidth]{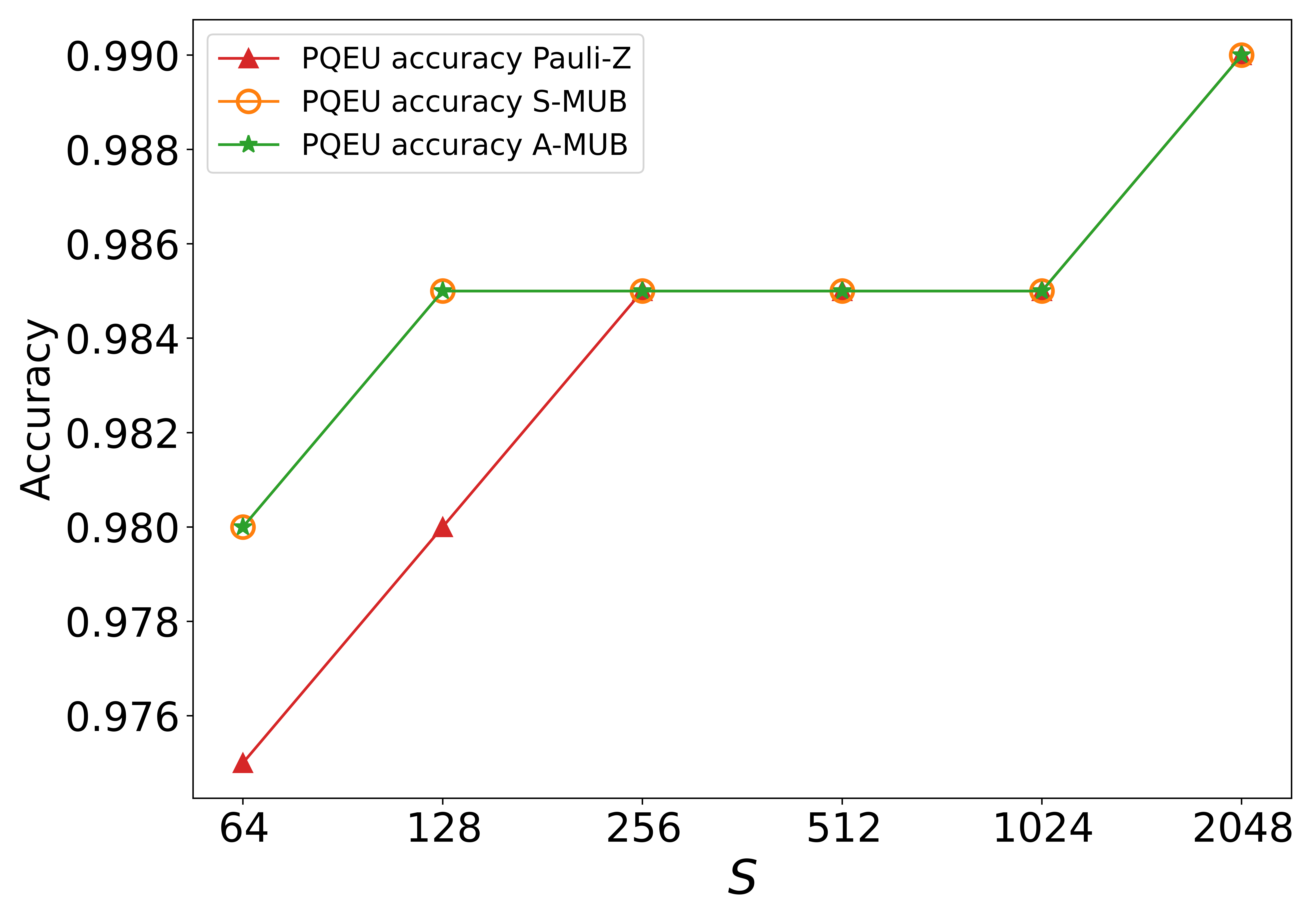}
		\caption{			Classification accuracy under finite-shot measurement noise.
		}
		\label{fig:shot_noise}
	\end{figure}

\textbf{Finite-shot measurement noise.}	On quantum devices, Pauli expectation values are estimated using a finite number of shots $S$.
	If $\mu = \langle \sigma \rangle$ denotes the ideal expectation value, then the empirical estimator
	\begin{equation}
		\hat{\mu}
		=
		\frac{1}{S}\sum_{i=1}^{S} x_i,
		\qquad x_i \in \{+1,-1\},
	\end{equation}
	satisfies
	\begin{equation}
		\mathbb{E}[\hat{\mu}] = \mu,
		\qquad
		\mathrm{Var}(\hat{\mu})
		=
		\frac{1 - \mu^2}{S}.
	\end{equation}
	For moderate $S$, we adopt the Gaussian approximation
	\begin{equation}
		\hat{\mu}
		\sim
		\mathcal{N}
		\left(
		\mu,
		\frac{1 - \mu^2}{S}
		\right),
	\end{equation}
	which enables differentiable simulation of shot noise.

In Fig.~\ref{fig:shot_noise}, we show the accuracy improves with the number of measurements increase which is reasonable since more measurements provides more information on the quantum states.	

	\begin{figure}[h]
		\centering
		\includegraphics[width=0.95\columnwidth]{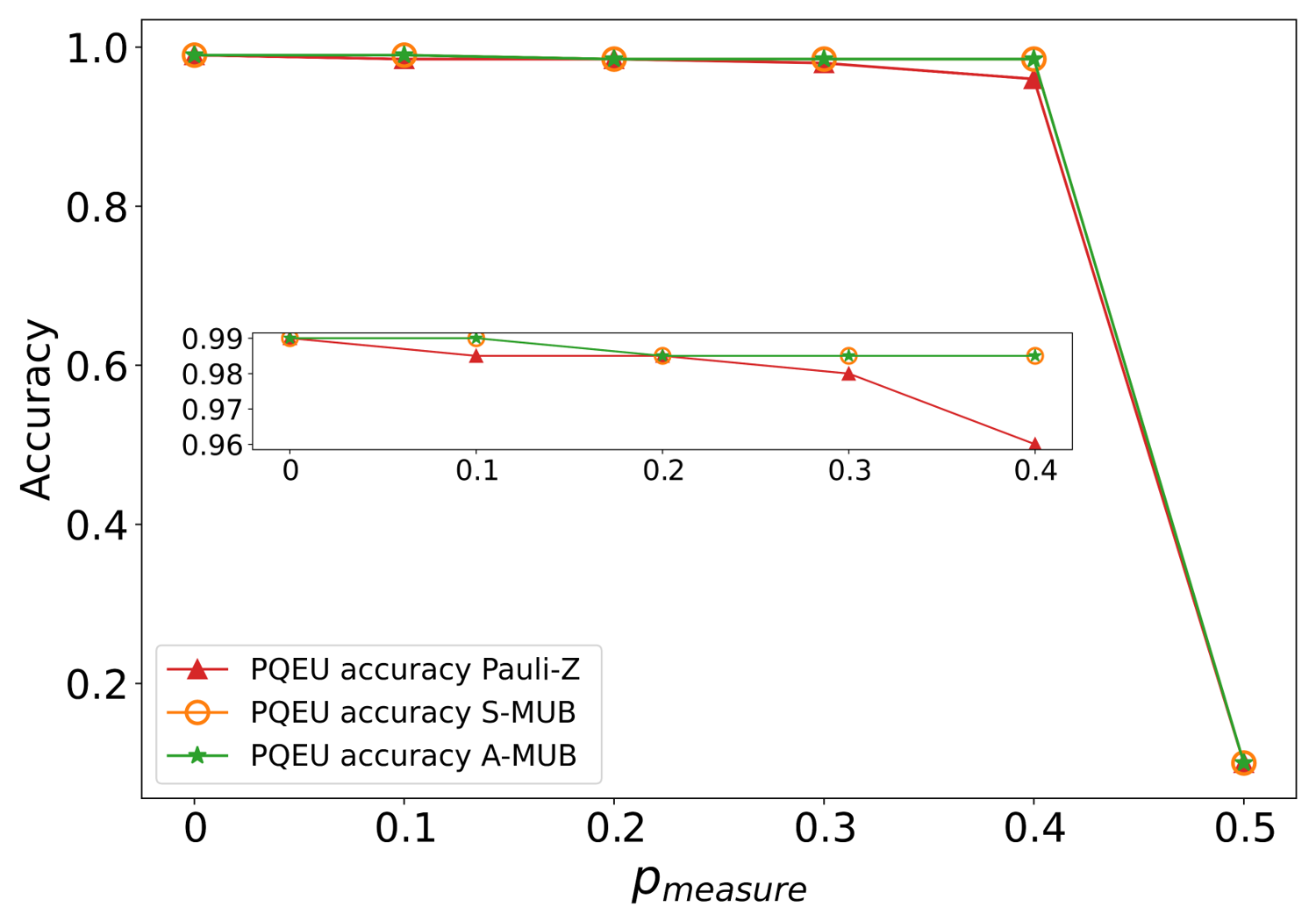}
		\caption{Classification accuracy under measurement readout noise.
			The parameter $p_{\mathrm{measure}}$ denotes the probability of classical bit-flip error applied to each measurement outcome.
			An inset shows a magnified view for $p_{\text{measure}} \le 0.4$.
			MUB-based strategies (S-MUB and A-MUB) demonstrate better robustness compared to single Pauli-Z measurement.
		}
		\label{fig:measure_noise}
	\end{figure}

\textbf{(3) Readout noise.} Symmetric readout bit-flip noise with probability $p_\text{measure}$ is added in our model as shown in Fig.~\ref{fig:measure_noise}. 
	We observe that while  $p_\text{measure}<0.5$, our model is able to compensate the bit flip readout, and ensure a high classification accuracy above $0.98$ for A-MUB strategy. While $p_\text{measure}=0.5$, corresponding to fully random measurement outcomes, the classification fails as a expected since there is no information in the measurement outcomes.

We study the time complexity of traditional QCNN and \change{PQEU} considering the following ingredients:  the number of rows $N$ and columns $M$ after gridding, the computational cost of a complete quantum stream  $T_{circ}$, including encoding, circuit evolution, and measurement.

For the traditional QCNN, each of the $N \times M$ image patches must be encoded and processed by an independent quantum circuit. Therefore, the total complexity is
\begin{equation}
    T_{\text{traditional}} = O(NM \cdot T_{circ}).
\end{equation}
In contrast, our \textbf{\change{PQEU}} approach  requires only parallel  executions on multi copies of the PQEU, the time required is independent of $N$ and $M$, but the numbe of the PQEU relies on the value of $N$ and $M$. The classical simulation time complexity in our GPU platform is
\begin{equation}
    T_{\text{\change{PQEU}}} = O(T_{circ}).
\end{equation}

Our analysis shows that \change{PQEU} reduces the time cost dependence on the number of patches and achieves an asymptotic speedup of $O(NM)$ over the traditional QCNN with the cost of multi compies of PQEU. In practice, this implies that as the size of input data grows, the computational burden of the conventional QCNN increases quadratically with the number of patches, whereas \change{PQEU} maintains a constant overhead. Such scalability makes \change{PQEU} particularly advantageous for large-scale or high-dimensional quantum-enhanced learning tasks.

We study the scalability of our approach by considering increasing the size of each input data (input resolution),  complexity of each data, and the total amount of data.

	\emph{Increasing input complexity.}
	When task complexity increases, the expressivity of the quantum kernel can be enhanced by moderately increasing the circuit depth, while keeping the qubit count. Since the PQEU functions as a localized convolutional kernel rather than a global variational circuit, increasing depth effectively improves feature extraction capacity without requiring additional qubits.

	\begin{table}[t]
		\centering
		\small
		\begin{tabular}{lcc}
			\hline\hline
			Method & MNIST & Fashion-MNIST \\
			\hline
			Quantum Deep Equilibrium Model\cite{yu2024qdeq} & 73.7\% & 72.1\% \\
			Resource-Efficient QCNN~\cite{res_qcnn2024} & 92.3\% & 88.5\% \\
			\hline
			\textbf{ParaQuanNet (Ours)} & 96.5\% & 84.3\% \\
			\hline\hline
		\end{tabular}
		\caption{Comparison with recent representative quantum neural network approaches. Reported values are taken from original papers under their respective settings.
			}
			\label{table:compare}
	\end{table}

	\emph{Increasing input resolution.}
	For larger images of size $H \times W$, the number of patches
	\[
	K = \frac{H}{p}\cdot\frac{W}{p}
	\]
	increases, but the quantum kernel circuit remains identical. Thus, qubit count and circuit depth per patch remain constant. The resource overhead scales linearly with the number of patches, similar to classical convolutional networks.

	\emph{Increasing the amount of data.}  When the dataset size increases, both the training and evaluation time scale approximately linearly with the number of samples. However, the computational complexity of a single forward inference remains unchanged, since the circuit structure and qubit count are fixed.

Overall, the scalability of our framework resembles classical convolutional architectures: dataset enlargement primarily increases patch training rather than circuit width, while increases in data complexity can be accommodated through moderate depth enhancement rather than drastic qubit expansion.

we also  compare with two recent representative quantum(-hybrid) image-classification works (2024--2025) to show the efficiency of our approach as shown in Table.~\ref{table:compare}
.

\subsection{experimental settings}

Our experimental model's hyperparameters were optimized through systematic grid searches and quantum hardware-aware tuning. Key configurations are shown in Table.~\ref{tab:parameters}. The loss function in our experiments is the  cross-entropy loss with $L_2$ regularization:  
\[
\mathcal{L} = -\sum_{c=1}^C y_c \log(\hat{y}_c).
\]

\begin{table}[htbp]
    \centering
    \begin{tabular}{cc}
      \hline\hline
            Parameter & Value\\
        \hline
            batch size & 32 \\
            optimizer & Adam\\
            learning rate & 0.002\\
            epochs & 40\\
            qubit &4\\
            Simulation Environment & TorchQuantum\\
        \hline\hline
    \end{tabular}
        \caption{Optimized parameter settings}
            \label{tab:parameters}
\end{table}

\begin{table*}[htbp]
\centering
\begin{tabular}{@{}lcccc@{}}
\hline
\textbf{Method} & \textbf{MNIST} & \textbf{Fashion-MNIST} & \textbf{EMNIST} & \textbf{CIFAR-10} \\ 
\hline\hline
QCNN Accuracy & 74.2\% & 73.8\% & 34.4\% & 33.1\% \\\hline
PQEU Accuracy & 93.1\% & 82.3\% & 78.3\% & 44.3\% \\\hline
PQEU AND S-MUB Accuracy & 96.1\% & 83.3\% & 78.6\% & 47.9\% \\\hline
PQEU AND A-MUB Accuracy & 96.5\% & 84.3\% & 78.6\% & 49.1\% \\\hline
Accuracy improvement & 22.3\% & 10.5\% & 44.2\% & 16\% \\
\hline\hline
\end{tabular}
\caption{Performance of the proposed \change{PQEU} model on benchmark datasets.}
\label{table:PQEU_QCNN}
\end{table*}

All our experiments are simulated in calssical GPU. Our experiments are implemented on a computer configured with two AMD EPYC 9554 CPUs, four NVIDIA A100 GPUs, 512GB of memory, a Ubuntu 20.04 system, Python 3.8.13, and the torchquantum 0.1.5 framework. Each experiment was verified five times. The seed value was set to [0,1,2,3,4] at the beginning of each experiment before the following operations: 
(1) dataset splitting, 
(2) model weight initialization, and 
(3) any stochastic data augmentation operations.

\subsection{Classification of classical data with our ParaQuanNet}
Our ParaQuanNet is also suitable for the classification tasks of classical data. To further validate the performance of the proposed ParaQuanNet model, we conducted additional experiments on several widely used benchmark datasets, including \textbf{MNIST}, \textbf{Fashion-MNIST}, \textbf{EMNIST}, and \textbf{CIFAR-10}. The classical data in Table.~\ref{table:PQEU_QCNN} are encoded into quantum states via the amplitude encoding method in Ref.~\cite{rath2024quantum}. Then the encoded quantum states are fed into our ParaQuanNet for the training and test processes. The result is shown in Table.~\ref{table:PQEU_QCNN}. Our ParaQuanNet dramatically improves the classification accuracies compared with the conventional QCNN. Specifically, for MNIST, the accuracies are improved by $22.3\%, 10.5\%, 44.2\%, 16\%$ for each of the testing data set. And we also witness the accuracy improvement with our proposed MUB measurement methods.

\hspace{2cm}
\section{Conclusion}
In this work, we identify the quantum generative circuits for the quantum data copyright protection in the coming quantum AI era. We propose a PQEU model for the parallel quantum date processing in quantum neural networks and propose two mutual unbiased basis measurement methods to effectively improve the performance of our parallel neural network. The high  identification accuracy of quantum generative circuits reveals that the various quantum DDPM processes inherit the different quantum space mappings, which is the key that our ParaQuanNet learns. Our approach is robustness to the data noise and the circuit level noise, which enable its applicaiton on NISQ devices. Our proposed method also applies to the classification tasks of classical data sets and the performance is much higher than conventional quantum neural networks. Our results demonstrate that ParaQuanNet offers a scalable and effective framework for the reliable identification of quantum circuits, enabling more accurate characterization of quantum generative models and strengthening the foundations for interpretable, verifiable, and ultimately  quantum machine intelligence.

\vspace{1cm}

\section*{Acknowledgements} \label{sec:acknowledgements}
  We acknowledge support from the National Natural Science Foundation of China~(Grant No.~62571434) and the  Fundamental Research Funds for the Central Universities.

\section*{Conflict of Interest}  
The authors declare no financial/commercial conflicts of interest. 

\bibliographystyle{unsrt}
\bibliography{QclassGmodel}

\begin{thebibliography}{10}

\bibitem{liao2017satellite}
Sheng-Kai Liao, Wen-Qi Cai, Wei-Yue Liu, Liang Zhang, Yang Li, Ji-Gang Ren, Juan Yin, Qi~Shen, Yuan Cao, Zheng-Ping Li, et~al.
\newblock Satellite-to-ground quantum key distribution.
\newblock {\em Nature}, 549(7670):43--47, 2017.

\bibitem{cacciapuoti2019quantum}
Angela~Sara Cacciapuoti, Marcello Caleffi, Francesco Tafuri, Francesco~Saverio Cataliotti, Stefano Gherardini, and Giuseppe Bianchi.
\newblock Quantum internet: Networking challenges in distributed quantum computing.
\newblock {\em IEEE Network}, 34(1):137--143, 2019.

\bibitem{li2025microsatellite}
Yang Li, Wen-Qi Cai, Ji-Gang Ren, Chao-Ze Wang, Meng Yang, Liang Zhang, Hui-Ying Wu, Liang Chang, Jin-Cai Wu, Biao Jin, et~al.
\newblock Microsatellite-based real-time quantum key distribution.
\newblock {\em Nature}, pages 1--8, 2025.

\bibitem{arute2019quantum}
Frank Arute, Kunal Arya, Ryan Babbush, Dave Bacon, Joseph~C Bardin, Rami Barends, Rupak Biswas, Sergio Boixo, Fernando~GSL Brandao, David~A Buell, et~al.
\newblock Quantum supremacy using a programmable superconducting processor.
\newblock {\em Nature}, 574(7779):505--510, 2019.

\bibitem{zhong2020quantum}
Han-Sen Zhong, Hui Wang, Yu-Hao Deng, Ming-Cheng Chen, Li-Chao Peng, Yi-Han Luo, Jian Qin, Dian Wu, Xing Ding, Yi~Hu, et~al.
\newblock Quantum computational advantage using photons.
\newblock {\em Science}, 370(6523):1460--1463, 2020.

\bibitem{krinner2022realizing}
Sebastian Krinner, Nathan Lacroix, Ants Remm, Agustin Di~Paolo, Elie Genois, Catherine Leroux, Christoph Hellings, Stefania Lazar, Francois Swiadek, Johannes Herrmann, et~al.
\newblock Realizing repeated quantum error correction in a distance-three surface code.
\newblock {\em Nature}, 605(7911):669--674, 2022.

\bibitem{gao2025signal}
Haoyang Gao, Leigh~S Martin, Lillian~B Hughes, Nathaniel~T Leitao, Piotr Put, Hengyun Zhou, Nazli~U Koyluoglu, Simon~A Meynell, Ania C~Bleszynski Jayich, Hongkun Park, et~al.
\newblock Signal amplification in a solid-state sensor through asymmetric many-body echo.
\newblock {\em Nature}, 646(8083):68--73, 2025.

\bibitem{o2016scalable}
Peter~JJ O’Malley, Ryan Babbush, Ian~D Kivlichan, Jonathan Romero, Jarrod~R McClean, Rami Barends, Julian Kelly, Pedram Roushan, Andrew Tranter, Nan Ding, et~al.
\newblock Scalable quantum simulation of molecular energies.
\newblock {\em Physical Review X}, 6(3):031007, 2016.

\bibitem{arguello2019analogue}
Javier Arg{\"u}ello-Luengo, Alejandro Gonz{\'a}lez-Tudela, Tao Shi, Peter Zoller, and J~Ignacio Cirac.
\newblock Analogue quantum chemistry simulation.
\newblock {\em Nature}, 574(7777):215--218, 2019.

\bibitem{PhysRevA.105.023513}
Cooper Doyle, Wei-Wei Zhang, Michelle Wang, Bryn~A. Bell, Stephen~D. Bartlett, and Andrea Blanco-Redondo.
\newblock Biphoton entanglement of topologically distinct modes.
\newblock {\em Phys. Rev. A}, 105:023513, Feb 2022.

\bibitem{king2025beyond}
Andrew~D King, Alberto Nocera, Marek~M Rams, Jacek Dziarmaga, Roeland Wiersema, William Bernoudy, Jack Raymond, Nitin Kaushal, Niclas Heinsdorf, Richard Harris, et~al.
\newblock Beyond-classical computation in quantum simulation.
\newblock {\em Science}, 388(6743):199--204, 2025.

\bibitem{PhysRevResearch.6.043163}
Wei-Wei Zhang, Zheping Wu, Hengyue Jia, Wei Zhao, Qingbing Ji, Wei Pan, and Haobin Shi.
\newblock Quantum versatility in pagerank.
\newblock {\em Phys. Rev. Res.}, 6:043163, Nov 2024.

\bibitem{xvng-rylk}
Wei-Wei Zhang, Chao Chen, and Jizhou Wu.
\newblock Self-induced manipulation of biphoton entanglement in topologically distinct modes.
\newblock {\em Phys. Rev. Appl.}, 24:054025, Nov 2025.

\bibitem{PhysRevLett.96.010401}
Vittorio Giovannetti, Seth Lloyd, and Lorenzo Maccone.
\newblock Quantum metrology.
\newblock {\em Phys. Rev. Lett.}, 96:010401, Jan 2006.

\bibitem{PhysRevLett.112.080801}
W.~D\"ur, M.~Skotiniotis, F.~Fr\"owis, and B.~Kraus.
\newblock Improved quantum metrology using quantum error correction.
\newblock {\em Phys. Rev. Lett.}, 112:080801, Feb 2014.

\bibitem{MONTENEGRO20251}
Victor Montenegro, Chiranjib Mukhopadhyay, Rozhin Yousefjani, Saubhik Sarkar, Utkarsh Mishra, Matteo~G.A. Paris, and Abolfazl Bayat.
\newblock Review: Quantum metrology and sensing with many-body systems.
\newblock {\em Physics Reports}, 1134:1--62, 2025.
\newblock Review: Quantum metrology and sensing with many-body systems.

\bibitem{Fadel_2025}
Matteo Fadel, Noah Roux, and Manuel Gessner.
\newblock Quantum metrology with a continuous-variable system.
\newblock {\em Reports on Progress in Physics}, 88(10):106001, oct 2025.

\bibitem{chen2023complexity}
Sitan Chen, Jordan Cotler, Hsin-Yuan Huang, and Jerry Li.
\newblock The complexity of nisq.
\newblock {\em Nature Communications}, 14(1):6001, 2023.

\bibitem{lau2022nisq}
Jonathan Wei~Zhong Lau, Kian~Hwee Lim, Harshank Shrotriya, and Leong~Chuan Kwek.
\newblock Nisq computing: where are we and where do we go?
\newblock {\em AAPPS bulletin}, 32(1):27, 2022.

\bibitem{ho2020denoising}
Jonathan Ho, Ajay Jain, and Pieter Abbeel.
\newblock Denoising diffusion probabilistic models.
\newblock {\em Advances in neural information processing systems}, 33:6840--6851, 2020.

\bibitem{sanchez2018inverse}
Benjamin Sanchez-Lengeling and Al{\'a}n Aspuru-Guzik.
\newblock Inverse molecular design using machine learning: Generative models for matter engineering.
\newblock {\em Science}, 361(6400):360--365, 2018.

\bibitem{goodfellow2020generative}
Ian Goodfellow, Jean Pouget-Abadie, Mehdi Mirza, Bing Xu, David Warde-Farley, Sherjil Ozair, Aaron Courville, and Yoshua Bengio.
\newblock Generative adversarial networks.
\newblock {\em Communications of the ACM}, 63(11):139--144, 2020.

\bibitem{wei2022emergent}
Jason Wei, Yi~Tay, Rishi Bommasani, Colin Raffel, Barret Zoph, Sebastian Borgeaud, Dani Yogatama, Maarten Bosma, Denny Zhou, Donald Metzler, et~al.
\newblock Emergent abilities of large language models.
\newblock {\em arXiv preprint arXiv:2206.07682}, 2022.

\bibitem{bowman2024eight}
Samuel~R Bowman.
\newblock Eight things to know about large language models.
\newblock {\em Critical AI}, 2(2), 2024.

\bibitem{bubeck2023sparks}
S{\'e}bastien Bubeck, Varun Chandrasekaran, Ronen Eldan, Johannes Gehrke, Eric Horvitz, Ece Kamar, Peter Lee, Yin~Tat Lee, Yuanzhi Li, Scott Lundberg, et~al.
\newblock Sparks of artificial general intelligence: Early experiments with gpt-4.
\newblock {\em arXiv preprint arXiv:2303.12712}, 2023.

\bibitem{sastry2024computing}
Girish Sastry, Lennart Heim, Haydn Belfield, Markus Anderljung, Miles Brundage, Julian Hazell, Cullen O'keefe, Gillian~K Hadfield, Richard Ngo, Konstantin Pilz, et~al.
\newblock Computing power and the governance of artificial intelligence.
\newblock {\em arXiv preprint arXiv:2402.08797}, 2024.

\bibitem{kandala2017hardware}
Abhinav Kandala, Antonio Mezzacapo, Kristan Temme, Maika Takita, Markus Brink, Jerry~M Chow, and Jay~M Gambetta.
\newblock Hardware-efficient variational quantum eigensolver for small molecules and quantum magnets.
\newblock {\em nature}, 549(7671):242--246, 2017.

\bibitem{cerezo2021variational}
Marco Cerezo, Andrew Arrasmith, Ryan Babbush, Simon~C Benjamin, Suguru Endo, Keisuke Fujii, Jarrod~R McClean, Kosuke Mitarai, Xiao Yuan, Lukasz Cincio, et~al.
\newblock Variational quantum algorithms.
\newblock {\em Nature Reviews Physics}, 3(9):625--644, 2021.

\bibitem{nghiem2021unified}
Nhat~A Nghiem, Samuel Yen-Chi Chen, and Tzu-Chieh Wei.
\newblock Unified framework for quantum classification.
\newblock {\em Physical Review Research}, 3(3):033056, 2021.

\bibitem{hur2022quantum}
Tak Hur, Leeseok Kim, and Daniel~K Park.
\newblock Quantum convolutional neural network for classical data classification.
\newblock {\em Quantum Machine Intelligence}, 4(1):3, 2022.

\bibitem{kolle2024quantum}
Michael K{\"o}lle, Gerhard Stenzel, Jonas Stein, Sebastian Zielinski, Bj{\"o}rn Ommer, and Claudia Linnhoff-Popien.
\newblock Quantum denoising diffusion models.
\newblock In {\em 2024 IEEE International Conference on Quantum Software (QSW)}, pages 88--98. IEEE, 2024.

\bibitem{wang2024quantum}
Ruhan Wang, Ye~Wang, Jing Liu, and Toshiaki Koike-Akino.
\newblock Quantum diffusion models for few-shot learning.
\newblock {\em arXiv preprint arXiv:2411.04217}, 2024.

\bibitem{minami2025generative}
Shunya Minami, Kouhei Nakaji, Yohichi Suzuki, Al{\'a}n Aspuru-Guzik, and Tadashi Kadowaki.
\newblock Generative quantum combinatorial optimization by means of a novel conditional generative quantum eigensolver.
\newblock {\em arXiv preprint arXiv:2501.16986}, 2025.

\bibitem{zhang2025quantum}
Wei-Wei Zhang, Xiaopeng Huang, Shenglin Shan, Wei Zhao, Beiya Yang, Wei Pan, and Haobin Shi.
\newblock Quantum data generation in a denoising model with multiscale entanglement renormalization network.
\newblock {\em Physica Scripta}, 100(6):065120, 2025.

\bibitem{shaik2025quantum}
Thanveer Shaik, Xiaohui Tao, Haoran Xie, and Robert Sang.
\newblock Quantum machine unlearning: Foundations, mechanisms, and taxonomy.
\newblock {\em arXiv preprint arXiv:2511.00406}, 2025.

\bibitem{parigi2024quantum}
Marco Parigi, Stefano Martina, and Filippo Caruso.
\newblock Quantum-noise-driven generative diffusion models.
\newblock {\em Advanced Quantum Technologies}, page 2300401, 2024.

\bibitem{zhang2024generative}
Bingzhi Zhang, Peng Xu, Xiaohui Chen, and Quntao Zhuang.
\newblock Generative quantum machine learning via denoising diffusion probabilistic models.
\newblock {\em Physical Review Letters}, 132(10):100602, 2024.

\bibitem{kwun2025mixed}
Gino Kwun, Bingzhi Zhang, and Quntao Zhuang.
\newblock Mixed-state quantum denoising diffusion probabilistic model.
\newblock {\em Physical Review A}, 111(3):032610, 2025.

\bibitem{he2026using}
Guang Ping Guang~Ping He.
\newblock Using a single circuit to compute the gradients with respect to all parameters of a quantum neural network: Using a single circuit to compute the gradients with respect to all parameters of a quantum neural network.
\newblock {\em Frontiers of Physics}, 21(8):083201, 2026.

\bibitem{liu2026output}
Yuxiang Liu, Fanxu Meng, Lu~Wang, Yi~Hu, Zaichen Zhang, and Xutao Yu.
\newblock Output prediction of quantum circuits based on graph neural networks: Output prediction of quantum circuits based on graph neural networks.
\newblock {\em Frontiers of Physics}, 21(6):063201, 2026.

\bibitem{cong2019quantum}
Iris Cong, Soonwon Choi, and Mikhail~D Lukin.
\newblock Quantum convolutional neural networks.
\newblock {\em Nature Physics}, 15(12):1273--1278, 2019.

\bibitem{wang2013multiple}
Yaohua Wang, Shuming Chen, Jianghua Wan, Jiayuan Meng, Kai Zhang, Wei Liu, and Xi~Ning.
\newblock A multiple simd, multiple data (msmd) architecture: Parallel execution of dynamic and static simd fragments.
\newblock In {\em 2013 IEEE 19th International Symposium on High Performance Computer Architecture (HPCA)}, pages 603--614. IEEE, 2013.

\bibitem{bent2015experimental}
Nicolas Bent, H~Qassim, AA~Tahir, D~Sych, Gerd Leuchs, Luis~Lorenzo S{\'a}nchez-Soto, E~Karimi, and RW~Boyd.
\newblock Experimental realization of quantum tomography of photonic qudits via symmetric informationally complete positive operator-valued measures.
\newblock {\em Physical Review X}, 5(4):041006, 2015.

\bibitem{zhang2018adaptive}
Aonan Zhang, Yujie Zhang, Feixiang Xu, Long Li, and Lijian Zhang.
\newblock Adaptive tomography of qubits: purity versus statistical fluctuations.
\newblock {\em Journal of Physics A: Mathematical and Theoretical}, 51(39):395304, 2018.

\bibitem{designolle2019quantifying}
S{\'e}bastien Designolle, Paul Skrzypczyk, Florian Fr{\"o}wis, and Nicolas Brunner.
\newblock Quantifying measurement incompatibility of mutually unbiased bases.
\newblock {\em Physical review letters}, 122(5):050402, 2019.

\bibitem{tavakoli2021mutually}
Armin Tavakoli, M{\'a}t{\'e} Farkas, Denis Rosset, Jean-Daniel Bancal, and Jedrzej Kaniewski.
\newblock Mutually unbiased bases and symmetric informationally complete measurements in bell experiments.
\newblock {\em Science advances}, 7(7):eabc3847, 2021.

\bibitem{farkas2023mutually}
M{\'a}t{\'e} Farkas, J{\k{e}}drzej Kaniewski, and Ashwin Nayak.
\newblock Mutually unbiased measurements, hadamard matrices, and superdense coding.
\newblock {\em IEEE Transactions on Information Theory}, 69(6):3814--3824, 2023.

\bibitem{yu2024qdeq}
Philipp Schleich, Marta Skreta, Lasse~Bj{\o}rn Kristensen, Rodrigo Vargas-Hernandez, and Alan Aspuru-Guzik.
\newblock Quantum deep equilibrium models.
\newblock {\em Advances in Neural Information Processing Systems}, 37:31940--31967, 2024.

\bibitem{res_qcnn2024}
Yanqi Song, Jing Li, Yusen Wu, Sujuan Qin, Qiaoyan Wen, and Fei Gao.
\newblock A resource-efficient quantum convolutional neural network.
\newblock {\em Frontiers in Physics}, 12:1362690, 2024.

\bibitem{rath2024quantum}
Minati Rath and Hema Date.
\newblock Quantum data encoding: A comparative analysis of classical-to-quantum mapping techniques and their impact on machine learning accuracy.
\newblock {\em EPJ Quantum Technology}, 11(1):72, 2024.

\end{thebibliography}

\end{document}